\documentclass[journal]{IEEEtran}

\usepackage{breqn}
\usepackage[english]{babel}
\usepackage[utf8]{inputenc}

\usepackage[short]{optidef}

\usepackage{amsmath}
\usepackage{amssymb}

\begin{document}

\title{Topology Learning Aided False Data Injection Attack without Prior Topology Information}

\author{
        Martin Higgins,
        ~Jiawei Zhang,
        Ning Zhang
        and Fei Teng

\thanks{

Martin Higgins and Fei Teng are with the Department of Electrical and Electronic Engineering, Imperial College London, London, SW7 2AZ, U.K. (Email:  f.teng@imperial.ac.uk).

Jiawei Zhang and Ning Zhang are with the State Key Laboratory of Power Systems, Department of Electrical Engineering, Tsinghua University, Beijing, China 
}


}

\markboth{IEEE Power \& Energy Society General Meeting 2021, Submitted on 06 November~2020}%
{Shell \MakeLowercase{\textit{et al.}}: Bare Demo of IEEEtran.cls for IEEE Journals}



\maketitle

\begin{abstract}
False Data Injection (FDI) attacks against power system state estimation are a growing concern for operators. Previously, most works on FDI attacks have been performed under the assumption of the attacker having full knowledge of the underlying system without clear justification. In this paper, we develop a topology-learning-aided FDI attack that allows stealthy cyber-attacks against AC power system state estimation without prior knowledge of system information. The attack combines topology learning technique, based only on branch and bus power flows, and attacker-side pseudo-residual assessment to perform stealthy FDI attacks with high confidence. This paper, for the first time, demonstrates how quickly the attacker can develop full-knowledge of the grid topology and parameters and validates the full knowledge assumptions in the previous work. 


\end{abstract}


\IEEEpeerreviewmaketitle

\section{Introduction}

\IEEEPARstart{C}{yber}-attacks against power systems have gained increased focus in recent years, with the events in Ukraine, Russia, Iran and Israel \cite{Liang2017TheAttacks} highlighting the importance of defence against cyber threats. Recently, a specific kind of attack against power system state estimation \cite{Monticelli1999StateApproach} has emerged, called the false data injection (FDI) attack \cite{Liu2011FalseGrids}. It has been shown that by altering these measurements in a very specific manner, negative consequences such as line overloading \cite{Tan2018Cyber-attackModel}, outage masking \cite {Liu2016MaskingAttacks} or load shedding \cite{Deng2017FalseSurvey}. Research into stealthy FDI has focused on both the development of new forms of attack and enhancing the detection of FDI.  This work focuses on enhancing attacks under limited knowledge assumptions. Below, we outline a literature review of relevant FDI attacks and the topology discovery technique which featured in our proposed attack.

\section{Background}

\subsection{FDI Attacks}

The principles behind FDI attacks were first outlined in \cite{Liu2011FalseGrids}. Liu \textit{et al}. used knowledge of the system topology to structure attack vectors so that the the system operator was unaware of the attack for a linearised DC model. However, in practice, it is unlikely that an attacker will have full knowledge of the underlying system topology. As a result, attacks that required a lower knowledge assumption were introduced, such as the incomplete knowledge attack in \cite{Rahman2012FalseGrids}, which showed that an DC system could be attacked with only partial knowledge of the system topology, and the 'blind' FDI attack, which showed a system could be attacked with no prior knowledge of the system topology. The original blind FDI attack required no prior system knowledge, provided the attacker had access to all meters within the attacked grid system \cite{Esmalifalak2011StealthGrid} \cite{Higgins2020StealthySystems}. The AC model has also been explored in the context of blind FDI attacks.  While the majority of these early works focused on a linear approximation of the FDI attack, it was shown in \cite{Hug2012VulnerabilityCyber-attacks} that an AC model was indeed possible, provided the attacker had knowledge of the system topology. In \cite{Chin2018BlindCommunications}, a geometric approach was applied for blind attacks against AC systems, but this attack still made some linearisation assumptions and lacked the significant flexibility in objective control granted by the full-knowledge attack. While papers such as \cite{Li2018AnalyzingOutages} offer a partial blind model that builds branch information. Other works in this field include \cite{Wang2020DetectionObserver}, where a method of FDI detection using unknown input internal observer is proposed. While in \cite{Kamal2020CyberattacksMeasures}, event-based triggers are used to enhance phasor measurement unit (PMU) based detection. Event-based FDI distributed detection is also explored in \cite{Higgins2020EnhancedDefence}. 

\subsection{Topology Discovery}
Attacks that can learn the underlying system topology offers more flexibility in how targets are chosen. There have been attempts to develop topology-discovery-style attacks, but they have been largely done under the assumptions of the linear model, such as in \cite{Li2018AnalyzingOutages}. In \cite{Bolognani2013IdentificationAnalysis},  voltage correlations are used to identify bus incidence, however, branch values are not calculated. This leaves a large portion of the required topology matrix unknown and, in practice, insufficient information for FDI attack. In other papers, such as, \cite{Wang2017PMU-BasedMatrix} a test is developed for estimating the dynamic Jacobin in the presence of topology changes, but this method requires PMU measurements, which are not always available for the attacker. Similarly, in \cite{Mousavi-Seyedi2015ParameterData} and \cite{Yu2018PaToPa:Grids}, models for network parameter estimation are suggested, but they also require PMU data in order to build an accurate model of the power system. As shown in \cite{Zhang2020TopologyMethod}, it is possible to evaluate network branch parameters without PMU data. An initial approximation can be made using regression via matrix operations; this gives a quick approximation of the per-unit network topology. This is then used as a starting point over which a fine identification is run. The fine identification uses a modified Newton-Raphson to get high-quality per-unit estimations of network topology. 

\subsection{Novel Contributions}

The topology learning technique is combined with attacker-side pseudo-residual assessment to create a topology-learning-aided FDI attack (TL-FDIA) that has the capabilities of a full knowledge attack with no prior system knowledge requirements. Our contributions are outlined as below:  
\begin{itemize}
    \item A topology-learning-aided FDI attack capable of attacking power systems under a blind assumption model (no branch or network incidence information available). The attack is committed against the AC power system and uses the latest state-of-the-art topology discovery techniques to build a model for the network.   
    \item We introduce an attacker-side criteria assessment via a pseudo-residual calculation to allow probabilistic assessment of attack success before any attack committed, allowing the attacker to ensure stealthiness. We also show regional pseudo-residuals can be used to verify local attacks even in the presence of global topology errors.
    \item We demonstrate how quickly the attacker can develop full knowledge of the system topology and parameters and effectively invalidate the full-system-knowledge assumptions in previous studies.    
 \end{itemize}

The rest of this paper is organized as follows; the problem formulation is outlined in Section 3;  Section 4 details the design of the FDI attack vectors; Section 5 introduces the basis of the topology learning algorithm; Section 6 contains the results of the TL-FDIA and Section 7 concludes the paper.

\section{AC State Estimation}

We consider a standard AC power system with real power flow measurements under the non-linear expression defined by

\begin{equation}
    P_{ij} = V_{i}^2 g_{ij} - V_{i}V_{j}g_{ij}\cos{\Delta \theta_{ij}}-V_{i}V_{j}b_{ij}\sin{\Delta \theta_{ij}}.
    \label{Pac}
\end{equation}

and reactive power flows by 
 
\begin{dmath}
    Q_{ij} = -V_{i}^2(b_{ij}+b_{ij}^{sh}) + V_{i}V_{j}g_{ij}\cos{\Delta \theta_{ij}} \\ -V_{i}V_{j}b_{ij}\sin{\Delta \theta_{ij}}.
    \label{Qac}
\end{dmath}

 $V$ and $\theta$ are the system states, while $P$ and $Q$ are the power measurements. This system is measured by estimating a set of $n$ state variables $\textbf{x} \in \mathbb{R}^{n\times1}$ estimated by analysing a set of $m$ meter measurements $\textbf{z} \in \mathbb{R}^{m\times1}$ and corresponding error vector $\textbf{e} \in \mathbb{R}^{m\times1}$ . The non-linear vector function $\textbf{h}(\textbf{.})$ relating meter measurements $\textbf{z}$ to states $\textbf{h}(\textbf{x}) = (h_1(\textbf{x}),h_2(\textbf{x}),...,h_m(\textbf{x}))^T$ is shown by 

\begin{equation}
    \textbf{z} = \textbf{h(x)} + \textbf{e}.
    \label{generalized state equation}
\end{equation}

The state estimation problem is to find the best fit estimate of $\hat{\textbf{x}}$ corresponding to the measured power flow values of $\textbf{z}$. Under the most widely used estimation approach, the state variables are determined by minimization of a WLS optimization problem as 

\begin{equation} \label{chisquaredfull}
 {\mathrm{min}_x}\,J(\textbf{x}) =  (\textbf{z}-\textbf{h(x)})^T\textbf{W}(\textbf{z}-\textbf{h(x)}).
\end{equation}

This is done using iterative processes, usually the Newton-Raphson \cite{Monticelli1999StateApproach} utilising the Jacobin $\textbf{J}$ of partial derivatives

$$
\textbf{J}=
\begin{vmatrix}
\frac{\delta h_1}{\delta x_1}&...&\frac{\delta h_1}{\delta x_n}\\
...&...&...\\
\frac{\delta h_1}{\delta x_m}&...&\frac{\delta h_m}{\delta x_n}\\
\end{vmatrix}.
$$

The aim with these iterative processes is to minimise the difference between the individual estimated values of power flows and the measured ones where the error (or line residual) $r_p$ for real powers is defined by

\begin{equation}
     e_p= -P_{ij}^m + V_{i}^2 g_{ij} - V_{i}V_{j}g_{ij}\cos{\Delta \theta_{ij}}-V_{i}V_{j}b_{ij}\sin{\Delta \theta_{ij}}.
    \label{Pac}
\end{equation}

with a similar equation for reactive power. At the system level, the error check for final decision making is based on the absolute value of the sum of errors known as the 2-norm difference between measured and estimated power flows defined by 

\begin{equation} \label{residual}
    r =  ||\textbf{z} -\textbf{h}\hat{\textbf{x}}||_2.
\end{equation}

We have an alarm value $\tau$, which is defined using engineering judgement, usually based on chi-squared testing criteria based on a 99\% confidence interval derived via regression of previous residual values, such that an alarm is raised if $r>\tau$.

\section{Topology-Learning-Aided FDI Attacks}
\subsection{Full-Knowledge AC FDI Attack }
If an attacker has knowledge of how the non-linear function is formed $\textbf{h}(\textbf{.})$ they can define a set of $\textbf{x}$ values to achieve their stated aims in terms of $P$ and $Q$ such that

\begin{equation} \label{residual}
    \textbf{z}_a =  \textbf{h}(\hat{\textbf{x}}+\textbf{c}).
\end{equation}

Where $\textbf{c}$ is an $n \times 1$ matrix denoting the desired bias injected into the system states (usually voltage angles) by the attacker and $\textbf{z}_a$ denotes the desired attack vector profile of measurements. The residual under such attack will therefore be defined by

\begin{equation} \label{residual}
    r =  ||\textbf{z}_a -\textbf{h}(\hat{\textbf{x}}+\textbf{c})||_2.
\end{equation}

The attacker can ensure this value is close to 0 as the injected (measured) value has been designed specifically to equal the one estimated using these flows. In practice, however, it is unlikely that an attacker will have the required knowledge an FDI attack, as this information will rarely be available publicly and maybe intentionally hidden.

\subsection{Attack Assumption and Overview of TL-FDIA }

 For TL-FIDA, we operate from the assumptions usually present in the blind attack models \cite{Yu2015BlindGrid}, as below:
\begin{itemize}
    \item The attacker has reading access to all measurements and can alter all or some real and reactive power measurements in the system.
    \item The attacker has no knowledge of system interconnection or branch admittance/resistance values. 
\end{itemize}

\begin{figure}
\centering
\includegraphics[width=2.5in]{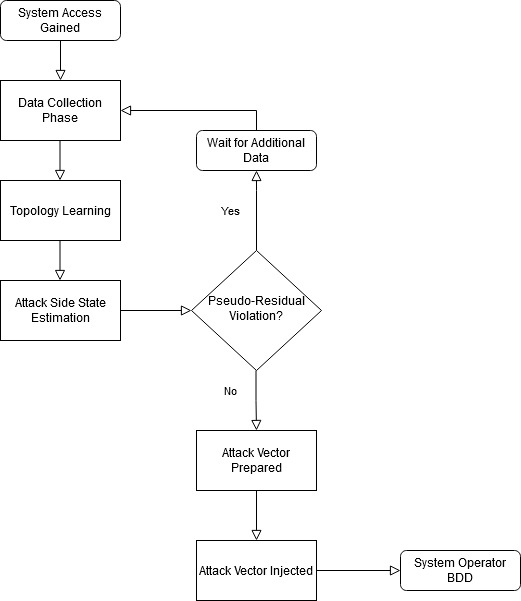}
\caption{Overview of the Topology Learning FDI Attack algorithm implementation.}
\label{fig:ACattackalgo}
\end{figure}

Keeping these assumptions in mind, the attacker will need to create a model of the power system only from  the available power flow measurements. Once the attacker has gained access to the system, the algorithm enters a period of data collection. When sufficient data has been received, the attacker attempts to perform the topology learning step of the attack based on the received data. This allows the attacker to subsequently perform an attack-side state estimation to verify the accuracy of the model using the derived topology. If the proposed vector passes the pseudo-state estimation residual check, the attacker can then proceed to attack phase. If not, the attacker waits for additional data and reruns the topology learning step of the attack. This proposed flow is outlined in Figure \ref{fig:ACattackalgo}. 

\subsection{Attack Side Verification}
Compared with the full knowledge attack, an important consideration for TL-FDIA is to know when they have collected enough data and are ready to attack. This can be difficult, as the attacker has zero prior information and no access to the system operator residual data, so any indication as to whether the proposed attack vector may pass BDD is based only on new inbound measurements. Consequently, we propose an attacker-side pseudo-residual calculation as an assessment on whether the attack can proceed based on     

\begin{equation} \label{residual}
    r_p =  ||\textbf{z}_a -\hat{\textbf{h}}(\hat{\textbf{x}}+\textbf{c})||_2.
\end{equation}

Where $\hat{\textbf{h}}$ is the estimated non-linear transformation function based on the estimated topology values and state measurements themselves.

\subsection{Sub-graph Residual}
In fact, even in the presence of global residual errors, the attacker may be able to identify subgraphs within the network where they can attack without altering other regions with poor residual performance. 

In practice, this will be similar to the incomplete-information-type attacks in \cite{Rahman2012FalseGrids}. Areas of high regional residual will be assumed to have incomplete knowledge, and other, lower-error regions can be attacked. Therefore, the attacker can use the regional meter error given by an alarm triggered, defined by    

\begin{equation} \label{residual}
    r_p^m =  \textbf{z}^m - \textbf{z}_{est}^m >\tau^m.
\end{equation}

With respect to the FDI attack, the sub-graphs are given by the number of non-zero terms in the column vector of the topology matrix $\textbf{H}$ for a given node $n$ or the Jacobian $\textbf{J}$ in the non-linear model. The attacker can then identify the corresponding sub-graphs related to this branch using the network incidence matrix $\textbf{I}$. For meter number m, 

\begin{equation}
 \textit{col}_n(\textbf{I}_m) =
    \begin{cases}
      1 & \text{meter $m$ is part of bus $n$ subgroup}\\
      0 & \text{meter $m$ is not part of the subgroup}.\\
    \end{cases}       
\end{equation}

The attacker then can structure the state adjustment vector $\textbf{c}$ such that $\textbf{I}_{n,m} = 0$. 

\section{Blind Topology Identification}
For the initial topology identification, we employ the method outlined in \cite{Zhang2020TopologyMethod}. The aim of topology identification is to identify the network incidence as well as the branch values for the conductance and susceptance matrices $[G_{ij}]$ \& $[B_{ij}]$. The technique we have employed here was originally used to map distribution networks for system operators.  However, we turn this method against the system operator for the purposes of an FDI attack where the attacker has limited system information. The method outlined by Zhang \textit{et al}. utilises a two-step identification process to identify per-unit branch topology information. The initial step introduces a regression to calculate approximations of conductance and susceptance. Initially, the regression uses a linearised approximation of the relationship between branch parameters, voltages, real and reactive powers to create a basis initial approximation. 

We can consider the matrix formulation for this in terms 

\begin{equation} \label{BIG}
    [P/V]= G_{ij}^{\#}  [V],
\end{equation}

\begin{equation} \label{BIG}
    [Q/V]= -B_{ij}^{\#} [V].
\end{equation}

$B_{ij}^{\#}$ \& $G_{ij}^{\#}$ are approximations of the real and imaginary branch elements. These are based on assumption of small state angle differences under the standard equations for real and reactive power injection. Given this approximation, the branch components can be extracted using matrix operations from which a solution for the approximate 

\begin{equation} \label{BIG}
    G_{ij}^{\#}= [P/V][V]^T([V][V]^T)^{-1},
\end{equation}

\begin{equation} \label{BIG}
    B_{ij}^{\#}= [Q/V][V]^T([V][V]^T)^{-1}.
\end{equation}

These initial steps are basic matrix operations. This means they can be performed quickly and with limited computational power. Under a DC approximation, these can on their own give reasonable approximations of the network branches incidences. Under the method proposed by Zhang \textit{et al}. they provide an initial approximation for the network topology which is used as the starting point for the fine identification stage. This is then followed by a modified Newton-Raphson, which incorporates the branch topology values to refine the approximation.  Given the power system bus injections under polar coordinates as 

\begin{equation}
\begin{bmatrix}\Delta p\\\Delta q \end{bmatrix}_{1\times2n} = \begin{bmatrix} \frac{\delta p}{\delta g}&\frac{\delta p}{\delta g}&\frac{\delta p}{\delta g}\\\frac{\delta p}{\delta g}&\frac{\delta p}{\delta g}&\frac{\delta p}{\delta g} \end{bmatrix}\cdot\begin{bmatrix} \Delta g\\\Delta b\\\Delta \theta \end{bmatrix}_{1\times(2m+n-1)}.
\end{equation}

where  $g$ \& $b$ are conductance and susceptance of $m$ branches. A pseudo-power flow calculation is performed and the generalised inverse is then applied to solve for the difference in both topology and state angle such that  

\begin{equation}
\begin{bmatrix} \Delta g\\\Delta b\\\Delta \theta \end{bmatrix} = \begin{bmatrix} \frac{\delta P}{\delta G}&\frac{\delta P}{\delta G}&\frac{\delta P}{\delta G}\\\frac{\delta P}{\delta G}&\frac{\delta P}{\delta G}&\frac{\delta P}{\delta G} \end{bmatrix}^+\cdot\begin{bmatrix}\Delta P\\\Delta Q \end{bmatrix}.
\end{equation}

This is used in the usual additive process to derive an estimation for the topology.  

\begin{equation}
\begin{bmatrix}  g\\ b\\ \theta \end{bmatrix} ^{(k+1)}= \begin{bmatrix}  g\\ b\\ \theta \end{bmatrix}^{k} \cdot \begin{bmatrix} \Delta g\\ \Delta b \\ \Delta \theta \end{bmatrix}.
\end{equation}

A full outline of the applied algorithm for the topology discovery component can be found in \cite{Zhang2020TopologyMethod}.

\section{Results \& Analysis}

This section assesses the performance of the proposed attack on IEEE-14 bus system. To replicate the real-time operation of a power system as closely as possible, system loads have been simulated using mock load profiles and the MATPOWER toolset \cite{Zimmerman2011MATPOWER:Education}. 

 

\subsection{Effectiveness of TL-FDIA}
\begin{figure}
\centering
\includegraphics[width=2.5in]{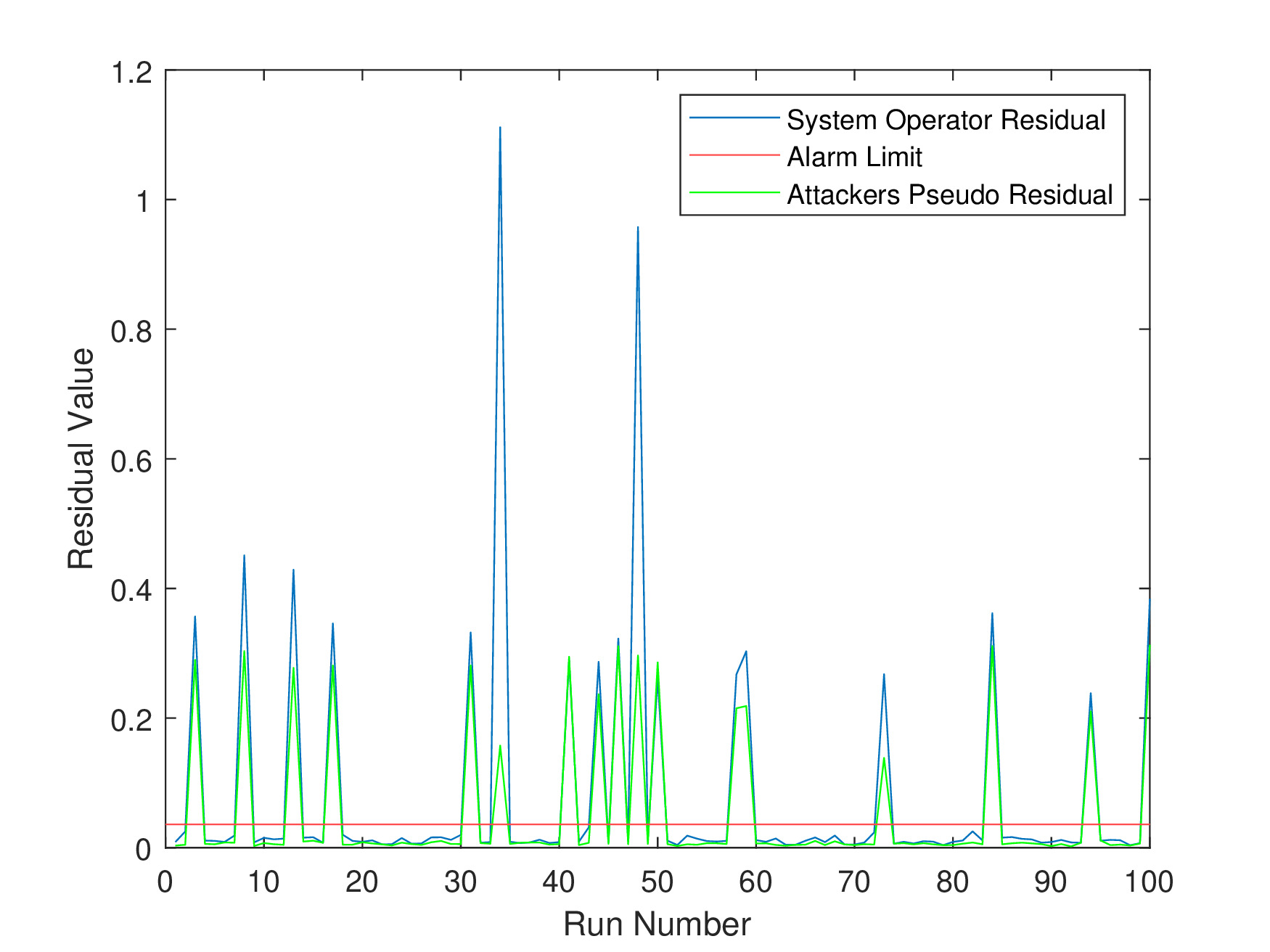}
\caption{Residual value measured by system operator and attacker in presence of TD-FDI attack against 1\% equivalent alarm. }
\label{fig:SOresidual}
\end{figure}

\begin{figure}
\centering
\includegraphics[width=2.5in]{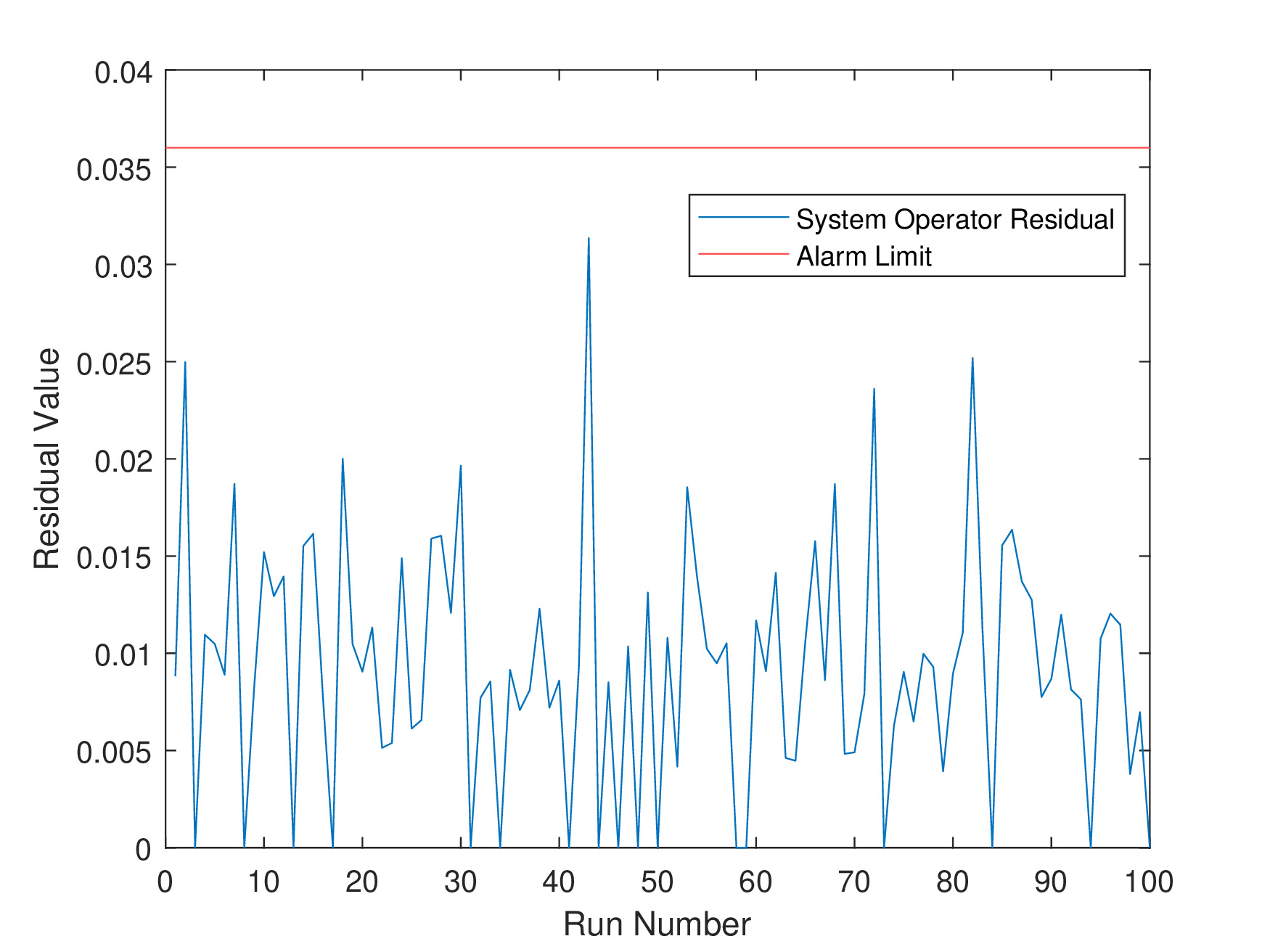}
\caption{Residual value measured by system operator TD-FDI attack with pseudo-residual decision statistic considered against 1\% equivalent alarm.}
\label{fig:residualwpseudocheck}
\end{figure}

In Figure \ref{fig:SOresidual}, we show the system residual as we apply the TL-FDIA using a voltage angle bias of around 15\% to bus 1. This results in power flow changes across buses 1-2 and 1-5. In blue we see the occasional residual spike past that of the acceptable alarm limit. In green we show the attackers pseudo-residual calculation, which, as expected, mimics the SO residual. As discussed, this pseudo-residual is used by the attacker as an attack-side assessment criterion. Without the pseudo residual we see around 80\% success in attack with 20 \% of values exceeding the SO residual. This is compared with an expected type-2 error of around 1\%. However, by applying the residual check we decrease this error significantly. In Figure \ref{fig:residualwpseudocheck}, we implement the pseudo-residual as a decision statistic, with the attacker choosing not to attack if they believe the residual will be violated. We note that the residual stays below the acceptable level and avoids detection. On timing, fine identification is achievable in around 5 seconds for a 33-bus system and 20 seconds for the 123-bus case. Increasing size is not a significant challenge for the algorithm as attacking region can always be divided into smaller regions to obtain quicker identifications.

\subsection{Data Requirement for TL-FDIA}
\begin{figure}
\centering
\includegraphics[width=2.5in]{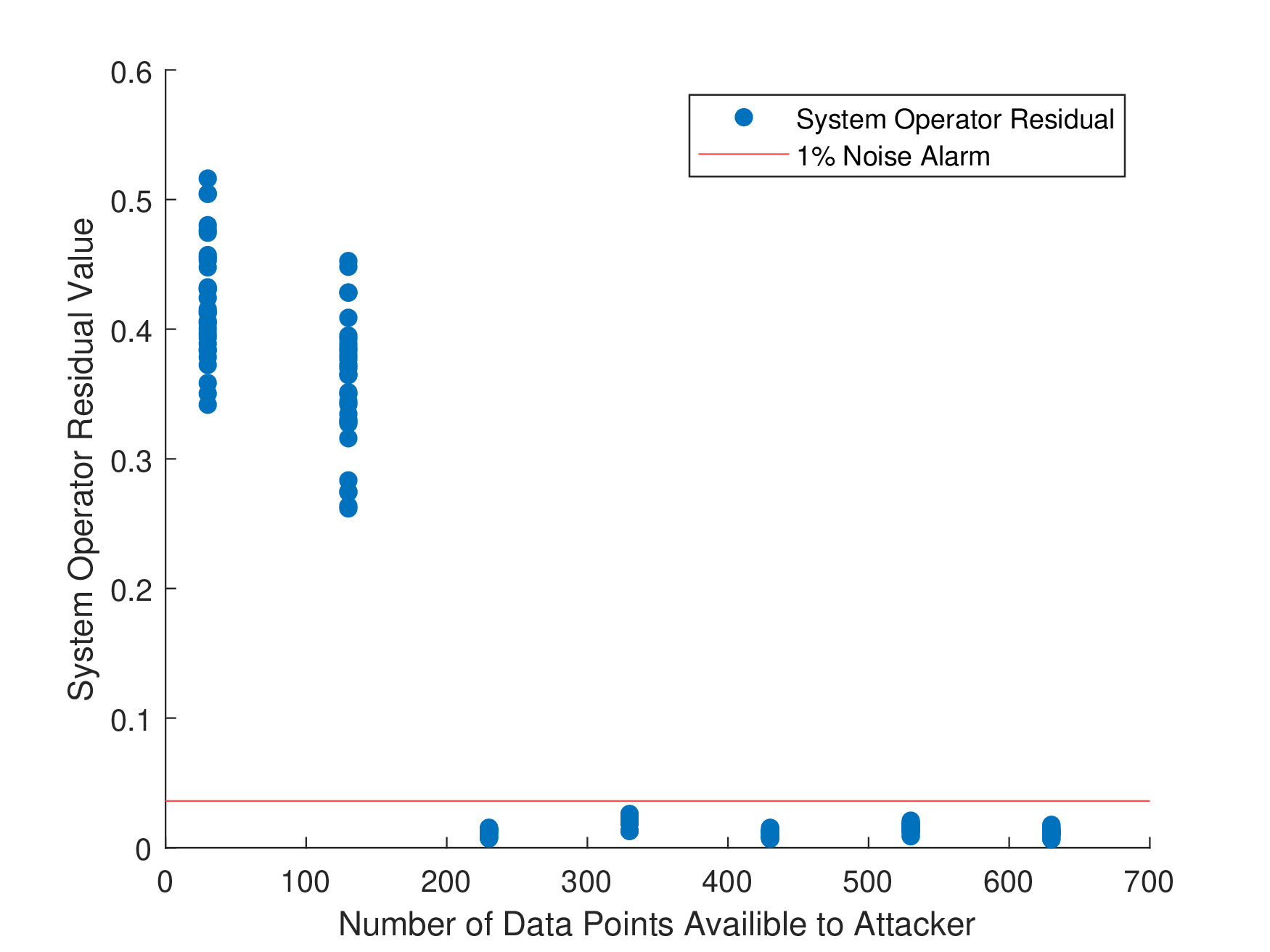}
\caption{Residual value measured by system operator in presence of TD-FDI attack with increasing number of available data points.}
\label{fig:SOMTDresidualincreasingruns}
\end{figure}

The previous successful attacks were performed with the assumption that 720 pieces of measurement data are available. However, the attacker may want to be ready to attack as soon as possible in order to avoid accidental exposure so it is critical to understand the minimum of measurement data that allows such an attack. In Figure \ref{fig:SOMTDresidualincreasingruns}, we show how pseudo-residual decreases with additional available data points. It is clear that with increasing amounts of measurement data, the system residual declines quickly. In the case of 1\% noise alarm and a 14-bus system, the attacker will likely wish to wait until at least 200 points are available before attempting to build a map of the system.

\begin{figure}
 \centering
 \includegraphics[width=2.5in]{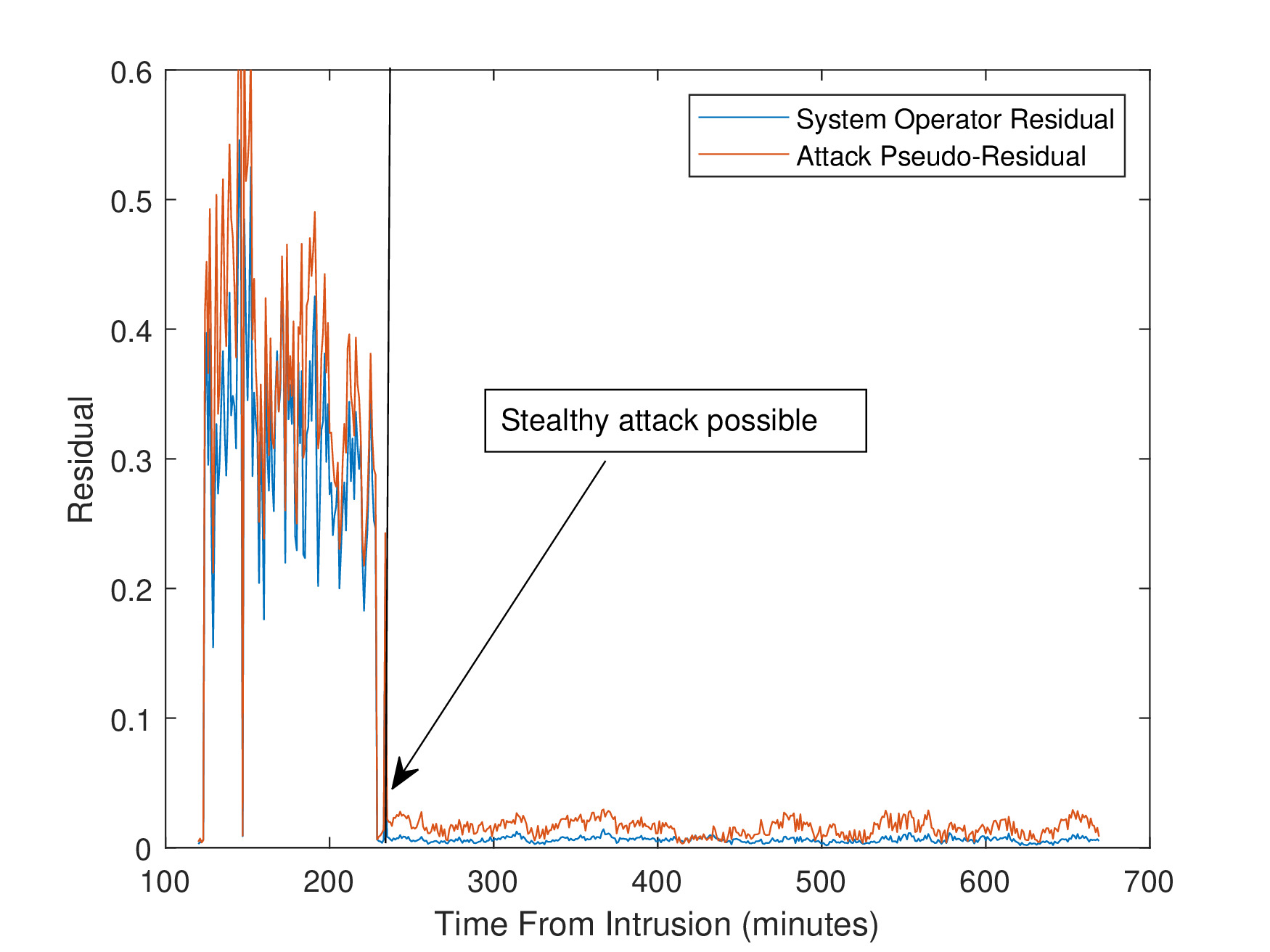}
 \caption{Residual value measured by system operator in presence of TD-FDI attack. X-axis shows number of minutes since intrusion.} 
 \label{fig:SOresidualTimedomain}
 \end{figure}

We simulate a time-domain attack scenario starting from the moment the attacker gains access to the full meter measurement to the moment the attack is ready to launch. The timing of attacks is a crucial consideration as the attacker seeks to minimise their time in the system.  In Figure \ref{fig:SOresidualTimedomain}, as individual measurements come in, we show how the system residuals from both operator and attacker change over time as additional measurements become available to the attacker. We assume the attacker has access to measurements in similar frequency to the central system operator, with state estimation and measurement set received once every minute. First, it is important to note again that, once the attacker's pseudo-residual converges to an acceptable level, the operator residual also approaches the level that will pass BDD, which demonstrates the effectiveness of attacker-side pseudo-residual assessment. In addition, the figure suggests that the attacker will need about 3-4 hours of data collection before they can initiate the attack without detection. Compared with the lengthy process of reconnaissance and penetrating the system, such a duration is almost negligible, which validates the full-knowledge assumptions in the previous works.   

\ifCLASSOPTIONcaptionsoff
  \newpage
\fi

\section{Conclusions and Future Work}


In this paper, we propose a topology-learning-aided FDI attack that combines topology learning techniques and attacker-side pseudo-residual assessment. We show via simulations on a 14-bus system that such an attack allows the performance of a full-knowledge AC FDI attack under blind assumptions. In the next stage, we intend to apply the algorithm in a larger and time-varying network to investigate its effectiveness. 



\bibliographystyle{IEEEtran.bst}
\bibliography{bare_jrnl.bbl}

\end{document}